\documentclass[pra, twocolumn, final, reprint, amsmath, amssymb, superscriptaddress, showpacs, a4paper, aps, 10pt, hidelinks, numbers, sort&compress, floats, balancelastpage]{revtex4-1}

\usepackage{graphicx, color} 
\usepackage[normalem]{ulem}          

\usepackage[colorlinks=true]{hyperref}
\hypersetup{pdfpagemode=None, linkcolor=black, citecolor=blue, urlcolor=blue}  

\usepackage[utf8]{inputenc}

\usepackage[normalem]{ulem} 
\usepackage{xcolor}

\newcommand{\1}{\ensuremath{\left|1 \right\rangle}}

\definecolor{britishracinggreen}{rgb}{0.0, 0.26, 0.15}
\definecolor{bulgarianrose}{rgb}{0.28, 0.02, 0.03}
\definecolor{darkred}{rgb}{0.90,0,0}
\definecolor{darkgreen}{rgb}{0,0.60,.2}
\definecolor{darkblue}{rgb}{0,0,1}
\definecolor{orange}{cmyk}{0,0.6,0.8,0}
\definecolor{lightblue}{rgb}{0.3,0.5,1}
\definecolor{lightgreen}{rgb}{0.4,0.80,.4}

\usepackage[utf8]{inputenc}
\usepackage{textcomp} 

\begin{document}

\title{Double-degenerate Fermi mixtures of $^6$Li and $^{53}$Cr atoms}


\author{A. Ciamei}
\email{ciamei@lens.unifi.it}
\affiliation{Istituto Nazionale di Ottica del Consiglio Nazionale delle Ricerche (CNR-INO), 50019 Sesto Fiorentino, Italy}
\affiliation{\mbox{European Laboratory for Non-Linear Spectroscopy (LENS), Universit\`{a} di Firenze, 50019 Sesto Fiorentino, Italy}}
\author{S. Finelli}
\affiliation{Istituto Nazionale di Ottica del Consiglio Nazionale delle Ricerche (CNR-INO), 50019 Sesto Fiorentino, Italy}
\affiliation{\mbox{European Laboratory for Non-Linear Spectroscopy (LENS), Universit\`{a} di Firenze, 50019 Sesto Fiorentino, Italy}}
\affiliation{Dipartimento di Fisica e Astronomia, Universit\`{a} di Firenze, 50019 Sesto Fiorentino, Italy}
\author{A. Cosco}
\affiliation{Dipartimento di Fisica e Astronomia, Universit\`{a} di Firenze, 50019 Sesto Fiorentino, Italy}
\author{M. Inguscio}
\affiliation{Istituto Nazionale di Ottica del Consiglio Nazionale delle Ricerche (CNR-INO), 50019 Sesto Fiorentino, Italy}
\affiliation{\mbox{European Laboratory for Non-Linear Spectroscopy (LENS), Universit\`{a} di Firenze, 50019 Sesto Fiorentino, Italy}}
\affiliation{Department of Engineering, Campus Bio-Medico University of Rome, 00128 Rome, Italy}
\author{A. Trenkwalder}
\affiliation{Istituto Nazionale di Ottica del Consiglio Nazionale delle Ricerche (CNR-INO), 50019 Sesto Fiorentino, Italy}
\affiliation{\mbox{European Laboratory for Non-Linear Spectroscopy (LENS), Universit\`{a} di Firenze, 50019 Sesto Fiorentino, Italy}}
\author{M. Zaccanti}
\affiliation{Istituto Nazionale di Ottica del Consiglio Nazionale delle Ricerche (CNR-INO), 50019 Sesto Fiorentino, Italy}
\affiliation{\mbox{European Laboratory for Non-Linear Spectroscopy (LENS), Universit\`{a} di Firenze, 50019 Sesto Fiorentino, Italy}}


\begin{abstract}
We report on the realization of a novel degenerate mixture of ultracold fermionic lithium and chromium atoms.
Based on an all-optical approach, with an overall duty-cycle of about 13 seconds, we produce large and degenerate samples of more than 2$\times 10^5$ $^6$Li atoms and $10^5$ $^{53}$Cr atoms, with both species exhibiting normalized temperatures of about $T/T_{F}$=0.25. 
Additionally, through the exploitation of a crossed bichromatic optical dipole trap, we can controllably vary the density and degree of degeneracy of the two components almost independently, and widely tune the lithium-to-chromium density ratio.
Our $^{6}$Li-$^{53}$Cr Fermi mixture opens the way to the investigation of a variety of exotic few- and many-body regimes of quantum matter, and it appears as an optimally-suited system to realize ultracold paramagnetic polar molecules, characterized by both electric and magnetic dipole moments. Ultimately, our strategy also provides an efficient pathway to produce dipolar Fermi gases, or spin-mixtures, of ultracold $^{53}$Cr atoms.  
\end{abstract}

\maketitle
\section{Introduction} 
\label{intro}

Quantum matter composed by unequal kinds of fermionic particles, such as quarks of different colors, or electrons belonging to different lattice bands, is known to promote a plethora of exotic phenomena \cite{Casabuoni2004,Andersen2011,Hammer2013,Yi2017,Wang2018,Jiang2021}, qualitatively distinct from those characterizing single-component systems. 
The combination of quantum statistics with a mass asymmetry, and a distinct response to external fields, of two different fermionic species, indeed provides an increased level of complexity, with a strong impact both at the few- and many-body level.
Heteronuclear mixtures of ultracold fermionic atoms, resonantly interacting close to a Feshbach resonance \cite{Chin2010}, have been already for many years regarded as clean and versatile frameworks optimally-suited for the exploration of novel quantum phases -- primarily in the context of unconventional superfluid pairing \cite{Liu2003,Forbes2005,Iskin2006,Parish2007,Baranov2008,Gubbels2009,Gezerlis2009,Baarsma2010,Gubbels2013,Wang2017,Pini2021} and quantum magnetism \cite{Keyserlingk2011,Sotnikov2012,Sotnikov2013,Massignan2013,Cui2013} -- and for the disclosure of exotic few-particle states \cite{Efimov1973,Kartavtsev2007,Nishida2008,Levinsen2009,Castin2010,Endo2011,Blume2012,Endo2012,Ngampruetikorn2013,Levinsen2013,Bazak2017,Bazak2017b, Liu2022A}, whose existence critically depends upon the mass asymmetry $M/m$ between the system constituents, and on the system dimensionality.

Specifically, mixtures that exhibit mass ratios 8.17$\lesssim M/m\!\lesssim$13.6 are particularly appealing, as they support, already in three dimensions, few-body clusters with $p$-wave orbital symmetry \cite{Kartavtsev2007,Endo2011,Endo2012,Blume2012A,Bazak2017}, completely unexplored thus far. Contrarily to Efimov states, these are expected to bear a universal character: Namely, their properties -- such as energy, wavefunction, etc. --  are entirely determined by the \textit{heavy-light} two-body scattering parameters. Analogously to Feshbach dimers, such non-Efimovian states are expected to feature a genuine \textit{halo} nature, with sizes on the order of the two-body scattering length $a$, and largely exceeding the  van der Waals range of the interatomic potential. Thus, their existence is not accompanied by increased inelastic decay processes \cite{Levinsen2011} typical of Efimov physics \cite{Naidon2017}, making them appealing not only from the few-body physics viewpoint, but also relevant from a many-body perspective. In fact, the existence of both fermionic trimers \cite{Kartavtsev2007,Endo2011,Endo2012} and bosonic tetramers \cite{Blume2012A,Bazak2017} -- composed by one \textit{heavy-light} dimer plus one or two additional heavy fermions, respectively -- may foster the emergence of  exotic normal and superfluid states, characterized by strong few-body correlations that add to, or may even overcome, the standard two-body ones. 

This is especially true for bi-atomic combinations with mass ratios close to the critical value $M/m\!\sim$8.17 \cite{Kartavtsev2007} ($M/m\!\sim$8.86 \cite{Bazak2017}),  
 for which a stable trimer (tetramer) state almost degenerate with the dimer-atom threshold exists. 
 In that case, weak deviations of the two-body interaction potentials from the zero-range limit -- encoded in a non-zero effective range parameter $R^*$ within the scattering amplitude \cite{Levinsen2011, Endo2012, Bazak2017} -- suffice to lift the few-particle cluster energies above the  threshold for their dissociation into a dimer plus excess heavy atoms. Namely, a non-zero value of $R^*/a$ plays qualitatively the same role of a decreased $M/m$, and its tuning allows one to turn the cluster state into a scattering resonance between heavy fermions and dimers. 
Notably, for cold atoms near a magnetic Feshbach resonance (FR), $R^*/a$ can be experimentally tuned by varying the magnetic field detuning from the FR pole.
This potentially offers the unprecedented possibility to resonantly control few-body \textit{elastic} interactions in experiments, and to explore how they affect the system properties at the many-particle level.
Yet, none of the Fermi-Fermi mixtures nowadays available, i.e. $^6$Li-$^{40}$K \cite{Wille2008,Voigt2009,Costa2010}, $^{40}$K-$^{161}$Dy \cite{Ravensbergen2018,Ravensbergen2020} and $^{6}$Li-$^{173}$Yb \cite{Hara2011,Green2020}, exhibits a mass asymmetry that allows to investigate such an appealing scenario, although related few-body effects, preempting the appearance of a trimer state, have been disclosed in Li-K \cite{Jag2014}.

Here, we report on the realization of a degenerate Fermi mixture made of $^6$Li alkali and $^{53}$Cr transition metal ultracold atoms. Characterized by a peculiar mass ratio $M/m\!\sim$8.8, and exhibiting various magnetic Feshbach resonances suitable for the  control of interspecies interaction \cite{Ciamei2022B}, such a bi-atomic combination appears as an uniquely-suited system for disclosing universal few-body clusters and related novel many-body regimes \cite{Endo2012,Bazak2017}, whose emergence could be fostered by tuning the $R^*/a$ ratio and the density imbalance, and by confining the mixture in a quasi-two dimensional geometry \cite{Levinsen2009, Liu2022A}.
Besides this major feature of the lithium-chromium system, in the regime of strong interatomic repulsive interactions, three-body recombination processes are predicted to be drastically suppressed for the specific Cr-Li mass asymmetry \cite{Petrov2003}, making this mixture a pristine platform for exploring Stoner's ferromagnetism \cite{Stoner1933} and related phenomena \cite{Jo2009,Scazza2017,Valtolina2017,Amico2018,Scazza2020}, ``immune'' to the pairing instability. 
Finally, recent \textit{ab initio} calculations \cite{Zaremba2022} foresee, for the ground state of the LiCr dimer, a sizable electric dipole moment of about 3.3 Debye, combined with a $S\!=\!5/2$ electronic spin, thereby making lithium-chromium mixtures also extremely appealing candidates for the realization of paramagnetic polar molecules.

Our strategy to produce degenerate lithium-chromium Fermi mixtures is formally similar to the all-optical one developed for the lithium-potassium system in the Innsbruck experiment \cite{Spiegelhalder2010}, and it consists in the following main steps: Realization of a cold Li-Cr mixture in a dual-species magneto-optical trap (MOT) \cite{Neri2020}; direct loading of the bi-atomic sample in an optical dipole trap; evaporative cooling of the lithium sample, populating the two lowest Zeeman atomic states and, simultaneously, sympathetic cooling of chromium atoms prepared in their lowest internal states.  
However, in spite of its conceptual simplicity, successful application of this approach to Li-Cr requires to tackle various challenges -- mostly connected with fermionic chromium, and its rather limited experimental investigation \cite{Chicireanu2006, Naylor2015, Neri2020}, especially when compared with the well-known $^6$Li system (see e.g. Ref. \cite{Inguscio2007-Book}). 
 
Specifically, three major issues make the production of ultracold chromium gases non-trivial. First, chromium suffers from rather strong light-assisted inelastic collisions \cite{Chicireanu2006}: This problem, only partially overcome in our previous work \cite{Neri2020}, limits the number of $^{53}$Cr atoms that could be collected in the MOT to roughly 10$^6$. 
Second, direct loading of chromium atoms from the MOT into a standard optical dipole trap realized with infrared laser beams has proved to be challenging \cite{Chicireanu2007,Beaufils2008}. This owes to an increased light-assisted inelastic collision rate \cite{Chicireanu2006,Bradley2000} connected with a high atomic density in the optical potential, and with the fact that laser light near 1 $\text{µm}$ effectively lowers the cooling transition frequency, shifting the detuning of the MOT light towards the blue of the relevant atomic line $^7S_3 \rightarrow ^7P_4$ at 425.5 nm \cite{Chicireanu2006, Neri2020}.       
Finally, while evaporative cooling of lithium mixtures is widely employed and well established, the efficiency of sympathetic cooling of chromium via Cr-Li collisions cannot be \textit{a priori} taken for granted. Our recent  studies of the $^6$Li-$^{53}$Cr collisional properties found a sizable, though not exceedingly high, interspecies scattering length of about 42 $a_0$ \cite{Ciamei2022B}, only slightly smaller than the one of Li-K \cite{Wille2008, Naik2011} and thus in principle sufficient  to guarantee a good Li-Cr thermalization rate, for long-enough evaporation ramps. 
In contrast with the Li-K case, however, which exhibits a potassium-to-lithium trap depth ratio of about 2  at 1070 nm \cite{Spiegelhalder2010}, the polarizability of chromium is about three times smaller than the lithium one at that wavelength. As such, a standard monochromatic infrared trap is not suited to limit the loss of chromium atoms during the evaporation stage.

In the following, we describe how we could overcome these challenges in the experiment also in relation to our previous work \cite{Neri2020}, and how we successfully obtain, with an overall duty cycle of less than 13 s, quantum degenerate samples comprising more than 2$\times\!10^5$ Li and $10^5$ Cr atoms, polarized in their lowest Zeeman states, at temperatures of about 200 nK, corresponding to $T/T_{F,Li}\!\sim\!T/T_{F,Cr}\!\sim$0.25 ($T_{F,i}\!=\hbar \bar{\omega}_i (6 N_i)^{1/3}/k_B$ denotes the Fermi temperature of a cloud of $N_i$ atoms, $i\!=Li, Cr$, in a harmonic potential with average frequency $\bar{\omega}_i$).

The paper is organized as it follows: In section \ref{MOT} we describe our new protocol to produce a dual-species Li-Cr MOT. In particular, we discuss how the $^{53}$Cr MOT atom number can be substantially increased with respect to previous studies \cite{Chicireanu2006, Neri2020}, reaching up to 8$\times\!10^7$ cold chromium samples within a 2 s loading time, not relying on magnetic trapping of $D$-state atoms \cite{Chicireanu2006} and in the presence of a large $^6$Li MOT of 10$^9$ atoms.
In section \ref{BODT} we present an efficient scheme to load simultaneously Li and Cr atoms in a bichromatic optical dipole trap (BODT) directly from the MOT. In particular, we describe a simple method to overcome the difficulties connected with the direct loading of chromium from a MOT into a IR trap, by realizing a ``dark-spot'' through a weak green beam at 532 nm, superimposed to the main trapping beam at 1073 nm \cite{Simonelli2019}.
In section \ref{Evap}, we characterize the evaporation trajectories followed by Li and Cr atoms, starting from about 250 $\text{µK}$ and reaching simultaneous quantum degeneracy at temperatures below 200 nK. In particular, we discuss how the exploitation of an interspecies narrow Feshbach resonance \cite{Ciamei2022B} enables us to substantially increase the sympathetic cooling efficiency at low temperatures,
towards the end of the evaporation ramp.
Finally, in section \ref{crossed} we describe how the implementation of a crossed bichromatic beam, superimposed to the main BODT trap, enables the controlled and independent tuning of the densities of the two atomic components, allowing us to strongly enhance the degree of degeneracy of chromium, and to reach highly-degenerate conditions for both Li and Cr species simultaneously.

\section{Dual species lithium-chromium MOT}
\label{MOT}

The experimental and optical setups employed for the present studies have been already described in our previous work \cite{Neri2020}. In the following, we briefly recall the lights needed to produce a cold Li-Cr mixture in a dual species MOT, referring the reader to Ref.~\cite{Neri2020} for more details.
For lithium, we follow the scheme developed in Ref.~\cite{Burchianti2014}. Laser cooling and trapping of such element is rather well-established, and it requires only two laser lights addressing the $D_2$ ($^2S_{1/2}\!\rightarrow\!^2P_{3/2}$) atomic line at 671 nm: the cooling light, addressing the $F=3/2 \rightarrow F'= 5/2$ transition, and the repumper ligth, detuned by 228 MHz from the cooling one, near resonant with the $F=1/2 \rightarrow F'= 3/2$ transition. 
With respect to the Li MOT performance we previously reported \cite{Neri2020}, a further optimization of the shaping of both MOT and Zeeman slower (ZS) beams allows us to increase the lithium atom number collected in the MOT from 4$\times\!10^8$ to $10^9$, after a typical loading time of 6 seconds.

For the $^{53}$Cr atomic component, a much more substantial increase in the MOT atom number is here achieved. 
As described in Refs.~\cite{Chicireanu2006,Neri2020}, laser cooling of fermionic chromium is based on the $^7S_3\!\rightarrow\!^7P_4$ atomic line. Besides the cooling light (addressing the $F_S = 9/2 \rightarrow F_P'= 11/2$ transition), three blue repumpers, resonant with the $F_S\!\rightarrow\!F_P'\!=\!F_S$+1 transitions with $F_S\!=7/2,5/2$ and $3/2$, are required to operate the MOT. 
Furthermore, even with all blue repumpers on, the MOT transition remains slightly leaky, since optically excited atoms can decay from the $^7P_4$ state onto underlying $^5D_{3}$ and $^5D_{4}$ metastable states. Therefore, three additional “red” repumpers are needed to fully close the cooling cycle \cite{Chicireanu2006}. Two lights near 663 nm (one light at 654 nm), pump atoms from the $F_D''=11/2$ and $F_D''=9/2$ hyperfine levels of the metastable $^5D_4$ states ($F_D''=9/2$ level of the $^5D_3$ state), back to the ground state via the $^7P_3$ electronic level.

While in our previous work \cite{Neri2020} only two red repumpers were implemented, here all three lights are exploited to operate the MOT, thus fully closing the cooling cycle.
Specifically, the addition of the second repumper at 663 nm, connecting the $F_D''=F_P'=9/2$ hyperfine manifolds of the $^5D_4\!\rightarrow\!^7P_3$ transition, leads to a 10$\%$ increase in the steady-state MOT atom number, approximately. 
After this, we have optimized the beam size of every stage to increase the atom number collected in the MOT. In particular, we have increased the MOT and repumper beam waists by about a factor of two relative to our previous setup \cite{Neri2020}, now featuring $1/e^2$ radii of about 0.65 cm, effectively increasing the capture volume by almost one order of magnitude.
Notably, these improvements on the chromium setup enabled us to identify a peculiar region in the detuning-intensity plane of the MOT cooling light, that we could not previously access \cite{Neri2020}, within which light-assisted losses are drastically suppressed, resulting in a 80-fold improvement in the $^{53}$Cr MOT atom number,  reaching up to $8 \times 10^7$ after a loading time of 2 seconds.

In order to understand our strategy and to interpret our experimental data, it is convenient to recall some textbook results about the loading dynamics in a MOT.
Quite generally, the atom number collected in a MOT follows a time evolution defined by the rate equation
\begin{equation}
\frac{dN}{dt}=\Gamma_L -\alpha N(t) - \frac{\beta}{\left\langle V \right\rangle} N^2(t),	
\label{rateeq}
\end{equation}
where $\Gamma_L$ is  the loading rate, $\alpha$ is a one-body decay rate, $\beta$ is the rate coefficient per unit volume for light-assisted collisions, and $\left\langle V \right\rangle$ denotes the (density-weighted) volume of the cloud. 
Since in our experiment we exploit all (blue and red) repumping lights, we can neglect the one-body loss term, and safely set $\alpha$=0.
Eq.~(\ref{rateeq}) then yields the asymptotic value for the collected atom number
\begin{equation}
N_{\infty}=\sqrt{\frac{\Gamma_L \left\langle V \right\rangle}{\beta}}.	
\label{Ninf}
\end{equation}
From Eq.~(\ref{Ninf}) one can immediately see that, in order to increase $N_{\infty}$, one ideally wants, besides maximizing the atomic flux $\Gamma_L$, (i) increase the MOT volume $\left\langle V \right\rangle$, and (ii) minimize the inelastic loss coefficient $\beta$.

Let us consider how these two quantities depend upon the MOT parameters, namely the (normalized) detuning $\delta/\Gamma$ and the saturation parameter $s_0=I/I_S$ (for the $^7S_3\!\rightarrow\!^7P_4$ chromium line, the natural linewidth is $\Gamma\!=2 \pi\!\times$5.02 MHz, and $I_{S}$=8.52 $\text{mW/cm}^2$ the associated saturation intensity).

In the limit of $s_0\!\ll$1 and $|\delta|\gg \Gamma$, one obtains that the MOT volume scales as \cite{metcalf1999}
\begin{equation}
\left\langle V \right\rangle \sim \left(\frac{\Gamma}{16 \mu' k_L} \right)^{3/2} \frac{(2 \delta/\Gamma)^6}{(b \, s_0)^{3/2}},	
\label{V}
\end{equation}
where $k_L$ denotes the laser wavevector, $b$ the magnetic-field gradient, and $\mu'$ the effective differential magnetic moment for the cooling transition.

The dominant light-assisted loss processes that affect a chromium MOT involve pairs of one ground $S$- and one excited $P$-state atom \cite{Chicireanu2006,Neri2020}. Thus,  denoting with $\Pi_P$ the $P$-state population, on quite general ground one expects the rate coefficient to scale as $\beta \propto \Pi_P (1-\Pi_P)$. Again considering the standard result for $\Pi_P$ for a two-level atom  \cite{metcalf1999},
\begin{equation}
\Pi_P = \frac{s_0/2}{1+s_0+(2 \delta/\Gamma)^2},	
\label{Ppop}
\end{equation}
and taking the low-intensity and large-detuning limit of Eq.~(\ref{Ppop}), one obtains that, up to a constant, 
\begin{equation}
\beta \sim s_0/(2 \delta/\Gamma)^2.	
\label{beta}
\end{equation}
Combining this dependence of $\beta$ upon the MOT parameters, with the one for $\left\langle V \right\rangle$ given by Eq.~(\ref{V}), one then expects the MOT atom number to feature a scaling of the kind
\begin{equation}
N_{\infty}\propto \frac{\sqrt{\Gamma_L}}{b^{3/4}}\frac{(\delta/\Gamma)^4}{s_0^{5/4}}.	
\label{Ninf_asym}
\end{equation}

From the overall trend of Eq.~(\ref{Ninf_asym}), one can see how, for a given loading rate $\Gamma_L$, light-assisted losses can be mitigated -- thereby substantially increasing $N_{\infty}$ -- by working at low $s_0$ values, large detunings, and weak field gradients of the MOT, although a compromise must be obviously found, in order to guarantee a sufficiently strong force and high capture velocity of the MOT.
\begin{figure}
\begin{center}
\includegraphics[width=\columnwidth]{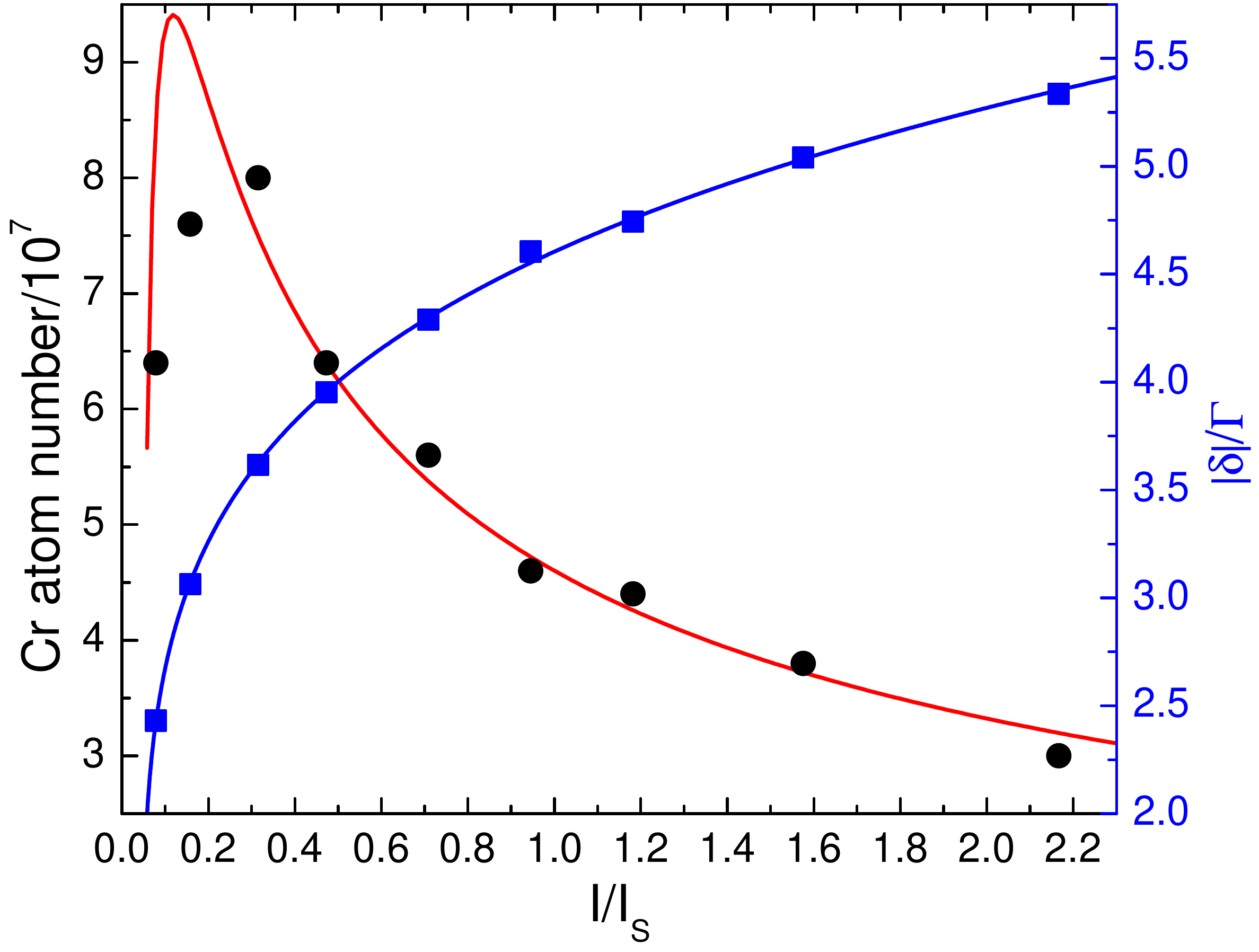}
\vspace*{-10pt}
\caption{Characterization of the maximum chromium atom number collected in the MOT after 2 s loading, as a function of $I/I_{S}$ (black circles, left axis). $I=2\times 2P/w^2$ is the peak intensity of one single (retro-reflected) MOT beam, characterized by a $1/e^2$ radius $w$=6.5 mm. Each data point is the average of at least five independent measurements. For each $I/I_S$, the experimental data exhibit a constant 10$\%$ uncertainty. 
For each value of $I/I_{S}$, the corresponding optimum detuning $|\delta|/\Gamma$, experimentally determined, is shown as blue squares, right axis. 
The blue line shows the best fit to a power law $|\delta|/\Gamma\!=\!(I/I_S)^\alpha$, see text.  
 The red solid line corresponds, up to a multiplicative factor, to $N_{\infty}$ given by Eq.~(\ref{Ninf_asym}), assuming the best-fitted power-law dependence of $|\delta|/\Gamma$ on $(I/I_S)$.}
\label{Fig1}
\end{center}
\vspace*{-10pt}
\end{figure}

A well-known system, where very strong light-assisted losses are successfully circumvented by following these concepts is metastable $^4$He$^*$ \cite{Pereira2001,Chang2014}: In that case, operating the MOT at large detunings on the order of $|\delta|\sim 40\Gamma$ while keeping large $s_0\!>$10 values to maintain a sufficiently high capture velocity, it is possible to collect more than 10$^9$ atoms within a few-second loading time.

For the $^{53}$Cr system, this strategy is challenging to follow. On the one hand, the saturation intensity (linewidth) of chromium is more than 50 times (3 times) larger than the one of He$^*$, and the limited amount of blue power available does not allow us to reach $s_0\gg$1, without diminishing the performance of transverse cooling and hyperfine pumping stages at the chromium oven \cite{Neri2020}, thus decreasing $\Gamma_L$. Moreover, $^{53}$Cr features a rich and rather dense hyperfine structure in both $S$ and $P$ states, absent in the case of metastable helium: large $s_0$ values can thus foster detrimental recycling effects \cite{Suominen1996,Weiner1999} that restore light-assisted collisions, and, combined with large detunings, may allow the cooling light to address undesired transitions, hence decreasing the MOT performances.
In light of these issues, and benefiting from a low ZS exit velocity, of about 15 m/s \cite{Neri2020}, we opt to follow a different strategy, based on minimizing $s_0$ while keeping relatively small light detunings of a few $\Gamma$.

Experimentally, we measure the atom number collected in the MOT after a 2 s loading time, exploring different (small) values of $s_0$. For each $s_0$ value, the light detuning is scanned until the maximum number is observed. The study is performed by intentionally keeping both $\Gamma_L$ and $b$ constant. Specifically, we work at a field gradient $b$=25 $\text{G/cm}$ along the vertical direction.
In order to better count the collected atoms, at the end of the loading stage we perform a compressed MOT (C-MOT) stage lasting about 6 ms, that strongly reduces the cloud size while not affecting the atom number. Then both MOT gradient and lights are turned off, and an absorption image is acquired after 2 ms time-of-flight. The atom number is then obtained through a 2D-Gaussian fit to the atomic sample.

Our results are summarized in Figure \ref{Fig1}. 
The atom number (black circles, left axis), is plotted as a function of the normalized single-beam peak intensity $I/I_S$, together with the corresponding optimum detuning experimentally identified (blue squares, right axis). One can notice how, throughout the scanned parameter space, a substantial increase in the collected Cr atoms is observed, relative to our previous study \cite{Neri2020}, and samples ranging from 30 to 80 million particles can be obtained. 
The behavior of $|\delta|/\Gamma$ versus $s_0$ is well fitted to a power-law, with exponent $\alpha$=0.22(1), see blue line. We note that such a value is relatively close to, but smaller than, the one that would maintain a constant $\beta/\left\langle V \right\rangle$ rate for light-assisted collisions. 
Based on the simple model discussed above, yielding Eqns.~(\ref{V}) and (\ref{beta}), this condition would indeed correspond to $\delta/\Gamma \propto s_0^{5/16}$, i.e. to $\alpha\!\sim$0.31. 

On the other hand, a non-constant loss rate is signaled by the observed variation of the MOT atom number, see black circles in Fig.~\ref{Fig1}. Remarkably, the observed behavior is nicely reproduced by the textbook model expectation Eq.~(\ref{Ninf}), shown as solid red line in Figure \ref{Fig1}, up to a multiplicative constant.   
The small mismatch between experiment and theory, visible at very low $s_0$ values, can be ascribed to the fact that in that regime the estimated MOT capture velocity becomes very close to, or even slightly smaller than, the exit velocity of our Zeeman slower. Parallel to that, the MOT size rapidly increases approaching the beam radius, thus making finite-size effects more important.

The identification of a region of MOT parameters able to strongly mitigate light-assisted losses allows us to greatly speed up and simplify the experimental routine to produce a large $^6$Li-$^{53}$Cr mixture in the cold regime. 
Since a loading time of 2 s suffices to reach $N_{\infty}$ for chromium, and that the MOT performances summarized in Figure \ref{Fig1} are not affected by the presence of an overlapping lithium cloud, there is no more need to pursue accumulation of Cr atoms in magnetically-trapped $D$ states \cite{Chicireanu2006}, a procedure that requires significantly longer loading times, and whose efficiency is limited by the presence of a large Li MOT \cite{Neri2020}.
Note also that working in the low $s_0$ limit substantially reduces the population of Cr atoms in the excited $^7P_4$ state, thus also strongly diminishing the leakage towards metastable $D$ states \cite{Chicireanu2006,Neri2020}. Indeed, when the MOT parameters are adjusted to the absolute optimum found in Fig.~\ref{Fig1}, the effect of all three red repumper lights on $N_{\infty}$ is much weaker, their complete switch-off resulting in only a two-fold reduction in the collected atom number. Furthermore, the fact that only a few mW of blue light suffice to realize the Cr MOT, the power of the transverse cooling and hyperfine-pumping beams at the Cr oven can be largely increased, thereby enhancing $\Gamma_L$. 

As a final remark, we note that while the  optimum loading conditions summarized in Fig.~\ref{Fig1} strongly reduce the Cr MOT density, they do not limit the capture efficiency of the C-MOT stage, operated at constant cooling light parameters. As a consequence, the strong increase in the MOT atom number directly turns into a significant density increase after the C-MOT, hence providing a substantial gain for the successive step of optical trap loading within our experimental routine. 

The ability to rapidly collect a large number of $^{53}$Cr atoms directly in the MOT allows for an optimized sequential loading of the Li-Cr mixture in our dual-species magneto-optical trap. The most convenient strategy we experimentally identified is summarized in the following: 
(i) We first load lithium atoms for about 5.9 s, at an optimum gradient of about $b$=45 $\text{G/cm}$ along the vertical direction. During this time, the chromium lights and ZS field are already on, although little Cr atom number is collected at this stage.
(ii) We switch off the Li Zeeman slower and decrease the MOT gradient down to $b$=25 $\text{G/cm}$, which is the optimum value found for chromium. The light detuning for lithium is correspondingly slightly diminished to ensure a good storage of this species during the chromium MOT loading. 
(iii) We operate the Cr MOT for about 2 seconds with the optimum light parameters reported in Fig.~\ref{Fig1}. 
(iv) We then turn off the Cr Zeeman slower (light and field), and adiabatically transfer the cold Li-Cr mixture from the quadrupole field of the MOT coils into that of a smaller set of ``Feshbach'' coils \cite{Neri2020}, yielding the same gradient but allowing for a faster switch off. 
(v) Finally, a 6 ms-long C-MOT phase is applied on both species simultaneously, in order to compress and cool the mixture. This is done, without changing the field gradient, by diminishing the intensity of the MOT lights, and moving the cooling frequency closer to resonance. Specifically, for chromium the C-MOT detuning is set to about -1.4 $\Gamma$ and the beams intensity is reduced to about 20$\%$, relative to that employed during the loading. For lithium, the detuning is moved from about -7 to -1.7 natural linewidths ($\Gamma_{Li}/(2 \pi)$= 5.87 MHz), and the light intensity is substantially reduced, passing from more than 17 $I/I_{S,Li}$ at the loading stage, down to about 0.5 $I/I_{S,Li}$ 
($I_{S,Li}$=2.54 $\text{mW/cm}^2$) \cite{Neri2020}.

At the end of this procedure, lasting 8 seconds overall, we obtain cold Li-Cr mixtures comprising 10$^9$ Li and 8$\times$10$^7$ Cr atoms, at a temperature of about 300 $\text{µK}$.

\section{Loading of lithium-chromium mixtures into a BODT}
\label{BODT} 
As anticipated in Section \ref{intro}, our experimental strategy is based on an all-optical approach conceptually analogous to the one employed for Li-K mixtures \cite{Spiegelhalder2010}. As a crucial step, this requires an efficient loading of the cold Cr-Li mixture, delivered by our dual species MOT discussed in Sec.~\ref{MOT}, into a high-power optical dipole trap.
However, also in this case a few factors make the Cr-Li system more challenging than the Li-K one. First, the chromium polarizability for standard infrared (IR) laser trapping lights at  1064 or 1070 nm 
 is only about 30$\%$ than the one for lithium, making the resulting IR trap not suited to guarantee an efficient sympathetic cooling of Cr. Indeed, a 1070 nm beam, characterized by power $P_{IR}$ (expressed in Watt) and $1/e^2$ waist $w_{IR}$ (in micron), yields a maximum trap depth for lithium (chromium) that, expressed in mK, is given by $U_{Li,IR}\sim-38.3 P_{IR}/w_{IR}^2$ ($U_{Cr,IR}\sim-12.7 P_{IR}/w_{IR}^2$). 

We mitigate this issue by superimposing a green beam at 532 nm to the IR trap. This second light is tightly confining for chromium, whereas it anti-confines lithium. Denoting the power and waist of such a second beam with $P_G$ and $w_G$, respectively, one finds in this case $U_{Li,G}\sim+39.2 P_{G}/w_{G}^2$ and $U_{Cr,G}\sim-23.5 P_{G}/w_{G}^2$. Therefore, by tuning the relative power of the two lights of this bichromatic optical dipole trap one can control the overall trap depth ratio for the two species.
Experimentally, the BODT is realized by overlapping our IR trap, already discussed in Ref. \cite{Simonelli2019} and based on a multimode fiber laser module from IPG Photonics (YLR-300) delivering up to 300 W, with a high-power laser at 532 nm. For the latter, we initially employed the GLR-50 module from IPG, able to produce up to 55 W. 
We later opted for a more reliable Sprout-G source by Lighthouse Photonics, nominally delivering up to 15 W.
The two BODT beams are recombined on a dichroic mirror
and then focused 
onto the center of the Li-Cr MOT clouds, 
with waists along the vertical (horizontal) direction of $w_{IR,z}$=44 $\text{µm}$ ($w_{IR,r}$=58 $\text{µm}$), and $w_{G,z}$=45 $\text{µm}$ ($w_{G,r}$=48 $\text{µm}$), for the IR and green light, respectively.

A second technical problem of Li-Cr is that 
the direct loading of atoms from the MOT into the optical trap, contrarily to the lithium case, see e.g.~Ref.~\cite{Burchianti2014}, has been found to be challenging
for chromium  \cite{Chicireanu2007,Beaufils2008,Bismut2011}. 
Besides increasing light-assisted losses owing to an increased density of the trapped cloud, the IR light shifts both $^7S_3$ and $^7P_4$ atomic levels -- connected by the main cooling transition -- to the red, with a shift for the excited state larger than the one of the ground state. 
Therefore, the detuning $|\delta|$ of the MOT light, experienced by Cr atoms within the IR trap, is effectively reduced (and it may eventually change sign). Light-shift measures, performed by monitoring the resonance frequency of absorption imaging of a cold Cr cloud in presence of our IR beam, yields a trap-averaged shift of -0.021(2) MHz$/$W. This implies that the (C-)MOT detuning, felt by atoms within the IR trap with $P_{IR}\!\sim$200 W, is moved towards resonance by about +$1\Gamma$. It thus becomes almost impossible to simultaneously guarantee a good efficiency of the C-MOT stage for Cr atoms both in- and out-side the IR trap, especially given the inhomogeneous intensity distribution of the trapping beam. One way to circumvent this problem is to flash the IR trap only once the MOT light is turned off. This, given our large MOT atom number and the high IR power at our disposal, allows us to capture about one million of $^{53}$Cr atoms within the IR beam at typical power of 130 W. Yet, this non-adiabatic loading method considerably heats up the sample, and it is far from being optimum also for the lithium component. 
More involved loading schemes, alternative to the instantaneous flash of the IR trap,
have been devised \cite{Chicireanu2007,Beaufils2008,Bismut2011}, which rely on the accumulation of metastable $D$-state atoms in a combined magnetic and optical potential.  

In our case, we found a convenient way,  offered by our BODT setup,  to successfully overcome this major technical issue.
The key point is that  the 532 nm light dramatically perturbs the cooling transition, owing to the presence of three atomic lines that connect the excited $^7P_4$ level to $^7D_5$, $^7D_4$ and $^7D_3$ states, respectively, all centered around 533  nm and featuring linewidths ranging from 0.9 to about 10 MHz. 
Thus the $^7S_3\!\rightarrow\!^7P_4$ line is strongly shifted towards higher frequencies already by a relatively weak laser field near 532 nm, blue-detuned from the $^7P_4\!\rightarrow\!^7D_{3,4,5}$ lines by less than one nanometer. 
Contrarily to the IR case discussed above, this implies that the effective detuning of the MOT light experienced by atoms within a 532 nm beam is strongly moved out of resonance.
Therefore, the green light of our BODT can be efficiently exploited to {(over-)}compensate the detrimental effect of the IR main beam on the Cr (C-)MOT, realizing an effective ``dark spot''.
\begin{figure}
\begin{center}
\includegraphics[width=\columnwidth]{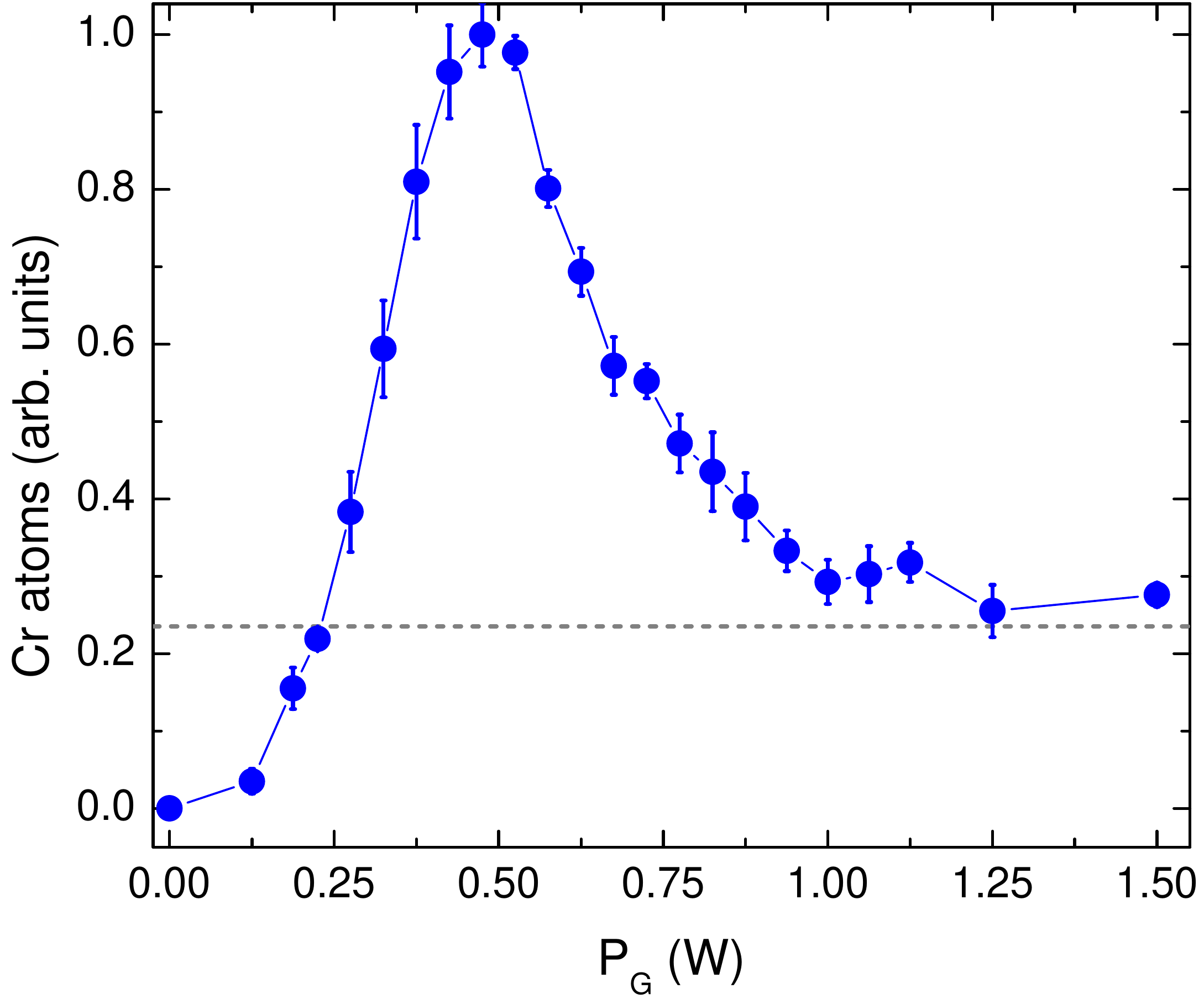}
\vspace*{-10pt}
\caption{Characterization of the chromium atom number collected in the BODT directly from the C-MOT, as a function of the green beam power acting as a ``dark spot''. The Cr population (blue circles), recorded in the BODT after a hold time of 100 ms through absorption imaging that follows a time-of-flight expansion of 200 $\text{µs}$, is normalized to the maximum value found throughout the scan, centered at 0.47(1) W. Each point corresponds to the average value of at least four independent measurements. Error bars represent the standard deviation of the mean.
For this data set, the IR beam has a fixed power of 130 W, and the parameters of the C-MOT stage are kept constant to their optimum values, obtained in the absence of the BODT. The green laser source employed for this study is the Sprout-G module by Lighthouse Photonics, featuring a wavelength of 532.2 nm.
}
\label{Fig2}
\end{center}
\vspace*{-10pt}
\end{figure}

We test the feasibility of this loading strategy by rising the IR beam up to 130 W  and, simultaneously, the green beam to a variable power level, through a 3 ms-long linear ramp that starts 1 ms before the chromium C-MOT stage. The laser cooling light parameters are kept fixed to the optimum values experimentally found in the absence of the optical trapping potential.  
5 ms after the end of the BODT ramp, both MOT lights and gradient are switched off. 
Turning off the cooling light 20 $\text{µs}$ before the repumper ones allows us to optically pump the Cr component into the $F_S\!=\!9/2$ hyperfine ground state.
After an additional hold time of 100 ms, we turn off the BODT and record, through an absorption image following a short time-of-flight expansion, the number of Cr atoms collected in the trap.
  
The result of this characterization is summarized in Fig.~\ref{Fig2}, that shows the chromium number loaded into the optical trap, normalized to its maximum value,
as a function of the power of the green beam (blue circles). 
One can notice how, without the green beam, almost no atoms are collected in the IR potential. By increasing the power of the 532 nm light, instead, we observe a sharp enhancement of the atom number which, for our specific BODT beam parameters, reaches its maximum at $P_{G}$=0.47(1) W. 
A further increase of the green beam power beyond this optimum value progressively diminishes the Cr atom number. For $P_G\!\geq$1.25 W, this approaches the value obtained by instantaneously flashing the IR trap right at the end of the C-MOT stage, marked by the horizontal gray line in Fig.~\ref{Fig2}. This behavior can be understood by considering that, once the green light reaches this power level, atoms falling within the BODT volume are effectively transparent to the C-MOT light, and thus completely unaffected by it. 

Owing to the strong inhomogeneity of the light shift experienced by the Cr C-MOT atoms throughout the BODT region at the loading, it is hard to quantify the actual light shifts based on the method employed to characterize the IR beam at the initial, high temperature. 
Measurements performed in the ultracold regime, where the Cr sample is much better localized near the center of the green laser, yields a peak shift of +38(5) MHz$/$W, characterized by a (positive) slope almost 2000 times larger than the IR (negative) one. Correspondingly, at the optimum value shown in Fig.~\ref{Fig2}, atoms residing at the center of the green beam experience an effective red shift of the cooling light of about -3.6 $\Gamma$. 

A quantitative analysis of the observed loading dynamics is quite involved and it goes beyond the scope of the present work. Anyway, we remark a few qualitative, general facts.
First, throughout the power range explored in Fig.~\ref{Fig2}, the green beam has negligible impact on the total trap depth, which is solely set by the high-power IR beam. Second, depending on the specific laser source employed for realizing the green BODT beam, the optimum power may quantitatively move to higher or lower values, but the qualitative trend remains unaffected, as long as the wavelength of the green light remains close to, but shorter than, 532.9 nm. Special care has to be taken to optimize both  collinearity and relative shapes of the IR and green waists for this method to optimally work: A relative axial displacement, and a too large or small $w_G$, relative to $w_{IR}$, can strongly reduce the performances of the ``dark spot''.
Third and most importantly, such a scheme leads to a substantial enhancement of the optical trapping efficiency, compared to the instantaneous flash of the IR beam, as it yields more than a four-fold improvement in the BODT atom number, and it does not cause any detectable excitation nor heating of the atomic sample. Under optimum conditions, this strategy allows us to store up to 4 $\times$ 10$^6$ Cr atoms in the optical trap, at temperatures of about 250 $\text{µK}$, slightly lower than the typical C-MOT one. Finally, we also remark that the absolute number of atoms that can be transferred in the BODT from the MOT with this strategy is found to scale linearly with the MOT atom number itself: Up to the largest MOT clouds of 8$\times\!10^7$ we can produce, we do not observe any saturation effect on the atomic samples in the optical trap, with an overall MOT-to-BODT constant transfer efficiency of about 5$\%$. 

Besides enabling to collect a significant amount of $^{53}$Cr atoms, which may be appealing also for single-species setups dealing with cold (fermionic or bosonic) chromium, this direct loading method is especially advantageous in our mixture experiment. Indeed, the presence of the weak green laser field is essentially irrelevant for the loading of the lithium component: up to 2$\times\!10^7$ $^6$Li atoms, with temperatures of about 280 $\text{µK}$ are stored in the BODT when the IR trap power is set to 130 W, with transfer efficiencies similar to those reported in Ref. \cite{Burchianti2014} for the single species case, although gray optical molasses based on the $D_1$ line are here not exploited \textit{before} loading Li in the BODT. 

Since the two species feature similar temperatures, the simultaneous loading of the Li-Cr mixture in the BODT does not perturb too strongly the chromium performance, although the initial trap depth ratio, uniquely set by the IR beam, yields at 130 W  $U_{Li,IR}\sim$1.9 mK and $U_{Cr,IR}\sim$0.65 mK, thus causing a rather strong asymmetry in the temperature-to-trap-depth ratio between the two components. In fact, while $\eta_{Li}\!=\!U_{Li,IR}/k_B T_{Li}\!\sim$7, for chromium we obtain $\eta_{Cr}\!=\!U_{Cr,IR}/k_B T_{Cr}\!\sim$3. For this reason, the chromium BODT population, after a hold time of 100 ms, in presence of the overlapping Li sample, is found to drop by almost a factor of 3.
This effect is partly reduced by applying a 350 $\text{µs}$-long $D_1$ molasses phase on lithium within the BODT \cite{Burchianti2014}, about 3 ms after the end of the C-MOT stage, once the magnetic field quadrupole gradient has been zeroed. This allows us to reduce the lithium temperature, although not substantially, from 300 down to approximately 220 $\text{µK}$, a value slightly lower than that of the Cr sample. At the end of the $D_1$ cooling, a 20 $\text{µs}$-long hyperfine pumping pulse is applied \cite{Burchianti2014}, which transfers all lithium atoms into the $F\!=1/2$ ground-state manifold. Finally, within 20 ms the green laser power is linearly ramped up to its maximum value, corresponding to a net power of 6 W  onto the atoms, leading to about a 10$\%$ increase (6$\%$ decrease) of the chromium (lithium) trap depth.

The application of the BODT loading method for $^{53}$Cr  discussed above, and its integration within our two-species experimental cycle, allows us to store in our optical dipole trap cold Li-Cr mixtures at about 250 $\mu$K, composed by 2$\times\!10^7$ $^6$Li atoms populating the two lowest Zeeman states $m_F\!=\!\pm\!1/2$ of the $F\!=\!1/2$ manifold, coexisting with about 2$\times\!10^6$ $^{53}$Cr atoms, asymmetrically distributed among the four lowest-lying Zeeman state of the $F\!=\!9/2$ hyperfine level. Specifically, without performing any Zeeman-selective optical pumping stage, we find that about 55$\%$ of the Cr sample is in the lowest Zeeman state, $m_F\!=\!-9/2$. The remaining Cr atoms are distributed among the three higher-lying levels, $m_F\!=\!-7/2$, $-5/2$ and $-3/2$, with relative populations of 25, 13 and 7$\%$, respectively. 
This represents our starting point for the successive stages of evaporative and sympathetic cooling, that we discuss in the following section.  
For convenience, in the following we denote the different Zeeman levels of both species with Li$|i\rangle$ and Cr$|i\rangle$ respectively, with $i\!=1,2,...$ labeling the atomic state starting from the lowest-energy one.

\section{Evaporative and sympathetic cooling stages}
\label{Evap}
Once the two species are loaded into the BODT, while the green BODT beam is ramped up to its maximum value, within 55 ms we also linearly increase the magnetic field bias up to 880 G, \textit{i.e.}  about 50 G above the broad Feshbach resonance occurring between the two lowest Zeeman states of lithium, Li$|1\rangle$-Li$|2\rangle$. At this field  \cite{Chin2010}, intra-species lithium collisions are  unitarity-limited at all temperatures relevant in this work, whereas inter-species Li-Cr collisions are at their background level, characterized by a scattering length $a_{bg}\!\sim$42 $a_0$ \cite{Ciamei2022B}. The magnetic-field curvature of our coils provides an additional in-plane harmonic confinement, characterized by a lithium (chromium) frequency of about 8.5 Hz (7.0 Hz), which adds to the BODT potential. 
The initial trap depth ratio between the two components, $U_{Li}/U_{Cr}\! \sim$3, and the comparably low initial value of $\eta_{Cr}\!\sim$3, are not optimal for an efficient storage of chromium atoms in the presence of the lithium sample. For this reason, we find experimentally convenient to start the evaporation immediately after the green BODT beam has been raised up. 

\begin{figure}[t!]
\begin{center}
\includegraphics[width=89mm]{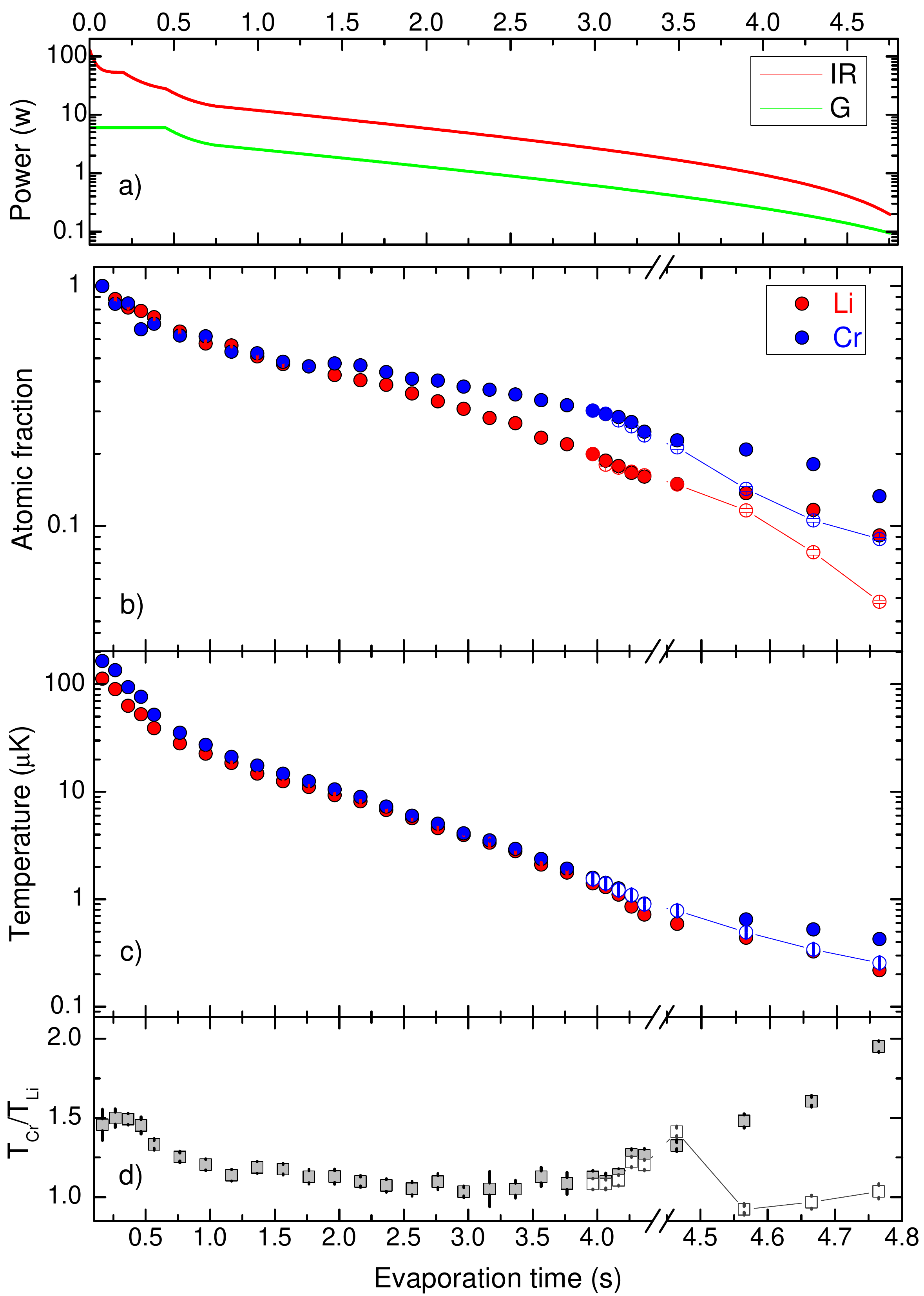}
\caption{
\textbf{(a)} Evolution of the IR and green BODT powers during the evaporation ramp. The IR power is reduced through four consecutive exponential ramps, lasting 0.2, 0.25, 0.3 and 4 s, respectively, and characterized by 1$/e$ decay times $\tau_1$=30 ms, $\tau_2$=125 ms, $\tau_3$=150 ms, and $\tau_4$=1.6 s.
The green power is decreased through two consecutive ramps, simultaneous to the last two IR ones, and featuring same durations and decay times.
\textbf{(b)} Evolution of the Li$|1\rangle$ (red circles) and Cr$|1\rangle$ (blue circles) atom number during the evaporation ramp. Both datasets are normalized to the atom numbers recorded after the first 165 ms of evaporation, where $N_{Li|1\rangle}$=5.1(1)$\times$ 10$^6$ and $N_{Cr|1\rangle}$=1.05(3)$\times$10$^6$. Empty symbols refer to the number evolution when the ``Feshbach cooling'' stage is applied, see text.
\textbf{(c)} Same as panel (b) but for Li and Cr temperatures.
For both (b) and (c) panels, numbers and temperatures are obtained from Gaussian fits to the atomic distributions, imaged after variable time-of-flight expansion. Each data is the average of at least three independent measurements, and error bars account for the standard deviation of the mean. Empty symbols refer to the temperature evolution when the ``Feshbach cooling'' stage is applied
\textbf{(d)} Ratio between chromium and lithium temperatures, without (full squares) and with (empty squares) application of the Feshbach cooling stage.}  
\label{Fig3}
\end{center}
\vspace*{-10pt}
\end{figure}

The evaporative cooling ramps, overall lasting for about 5 s, are performed by decreasing the power of the BODT beams, hence the trap depth, through a series of exponential ramps, shown in Fig.~\ref{Fig3}(a) for the IR and green lights, respectively. 
Fig.~\ref{Fig3}(b) and (c) show the corresponding evolution of the normalized atom number and temperature, for the Li$|1\rangle$ (red circles) and Cr$|1\rangle$ (blue circles) component, respectively, extracted from Gaussian fits to the density distributions, monitored via spin-selective absorption imaging following time-of-flight expansion. 
The Li$|2\rangle$ sample, not shown, throughout the evaporation stage is found at a temperature equal to the one of Li$|1\rangle$, and the corresponding atom number, relative to that of Li$|1\rangle$, is roughly constant at a value $N_{Li|2\rangle}/N_{Li|1\rangle}$=0.71(5).
Additionally, Fig.~\ref{Fig3}(d) displays the ratio between the chromium and lithium temperatures throughout the evaporation stage.

During the first 400 ms, evaporative cooling of lithium is established by decreasing only the IR beam power, from 130 W down to 28 W. This first step, sufficiently slow to allow for intra-species thermalization of lithium, is somewhat too fast for the chromium component, the temperature of which is found to be about 50$\%$ higher than the lithium one, see Fig.~\ref{Fig3}(d). Despite the rather poor efficiency of sympathetic cooling observed within this initial stage, such a ramp allows us to rapidly diminish the $U_{Li}/U_{Cr}$ ratio, from the initial value of 3, down to about 1. 
Such a ratio, which we keep roughly constant during the following evaporation stages, could in principle be further decreased keeping the overall Cr (or Li) BODT depth constant while increasing the green power relative to the IR one. However,  we have found that, owing to the slight mismatch in the two beam sizes, a too large green power strongly perturbs the BODT potential experienced by the lithium component, decreasing the evaporative cooling efficiency, and it reduces the Li-Cr density overlap, thus the efficiency of sympathetic cooling of chromium.

Besides carefully tuning the power of the two BODT lights, in order to maintain the chromium cloud well overlapped to the lithium one at all times, we also minimize the differential gravitational sag of the two components by applying a magnetic-field gradient $b$ along the vertical direction to counterbalance the gravitational force. Experimentally, we find an optimum value of about 1.6 G/cm, which corresponds to an almost perfect levitation of the chromium component, and to an effective weak ``anti-gravity'' for lithium, of about -$g/2$. 

For evaporation times longer than 0.5 seconds, where $U_{Li}/U_{Cr}\!\sim\!$1, the observed trajectories signal a good inter-species thermalization and a satisfactory sympathetic cooling. 
 The observed decrease in atom number is indeed significantly smaller for the Cr than for the Li component (see Fig.~\ref{Fig3}(b)), while the chromium temperature closely follows the lithium one with less than 15$\%$ mismatch, see Fig.~\ref{Fig3}(c) and (d), up to about 4 s. Here, we obtain about 3$\times$10$^5$ Cr$|1\rangle$ atoms at $T_{Cr}\!\sim\!$1.5 $\text{µK}$, coexisting with about 1.1$\times$10$^6$ Li$|1\rangle$ and 7.5$\times$10$^5$ Li$|2\rangle$ atoms at $T_{Li}\!\sim$1.35 $\text{µK}$, close to the onset of quantum degeneracy for the two lithium components.
By further decreasing the BODT trap depth following the trajectories shown in Fig.~\ref{Fig3}(a), we observe a progressive increase of $T_{Cr}/T_{Li}$, see Fig.~\ref{Fig3}(d). 

We initially tried to overcome such a limited inter-species thermalization by significantly extending the evaporation ramps in this final stage. Although application of slower ramps indeed help the two components to thermalize, we found that the increased duration of the evaporation routine strongly decreased the lithium atom number and phase-space density in the BODT. 
An alternative and much more convenient way to circumvent this problem is offered by the presence of various $s$-wave Li-Cr Feshbach resonances, located at fields above 1400 G \cite{Ciamei2022B}. In particular, the Li$|1\rangle$-Cr$|1\rangle$ mixture possesses a $\sim$0.5 G-wide Feshbach resonance at 1414 G, and the Li$|2\rangle$-Cr$|1\rangle$ combination exhibits a resonance of similar character around 1461 G.
Both features are immune to two-body losses \cite{Ciamei2022B} and, in spite of their relatively narrow character, allow us to magnetically control the Li-Cr scattering length $a$, and thus to increase the Li-Cr elastic scattering cross section well above its background value.

In order to exploit such a possibility, about 1.5 s after the start of the evaporation stage, we linearly ramp the magnetic field from 880 G, up to 2 G above the center of one of either resonances. There, the Li-Cr scattering length is not significantly different from its background value, $a\!\sim\!a_{bg}$, and also the intra-species Li$|1\rangle$-Li$|2\rangle$ scattering length approaches its large and negative background value, of about -2500 $a_0$ \cite{Zurn2013}.
About 4 s after the start of the evaporation, we then reduce the magnetic-field detuning to $\lesssim$100 mG from the resonance center, correspondingly tuning the Li-Cr scattering length to $a\!\lesssim$-200 $a_0$, yet not causing a significant enhancement of inter-species three-body losses.

While a detailed characterization of such a ``Feshbach cooling'' mechanism near a narrow resonance will be subject of a future study, the empty symbols in Figure \ref{Fig3}(b)-(d) panels highlight its impact on the final part of the evaporation ramps. One can see how, for fixed BODT power ramps, an increased Li-Cr scattering rate negligibly affects the Li temperature, whereas it causes a strong decrease of the Cr one, allowing us to perfectly cancel the relative temperature mismatch, see empty squares in Fig.~\ref{Fig3}d. The much quicker inter-species thermalization is accompanied by a more sizable atom loss of both species, see blue (red) empty circles in Fig.~\ref{Fig3}b for the Cr (Li) component. 
Yet, this only moderately decreases the degree of degeneracy of lithium, while for chromium the atom loss is outweighed by the strong temperature reduction, resulting in a substantial increase in the Cr phase-space density.
By following this protocol, overall lasting less than 5 s, we are able to produce degenerate lithium-chromium Fermi mixtures slightly below 200 nK, comprising up to 3.5$\times$10$^5$ Li$|1\rangle$ (2.5$\times$10$^5$ Li$|2\rangle$) atoms at $T/T_{F,Li}\!\sim$0.17(1) ($T/T_{F,Li}\!\sim$0.19(1)), coexisting with about 10$^5$ Cr$|1\rangle$ atoms at $T/T_{F,Cr}\!\sim$0.45(7). 
 The corresponding degree of degeneracy $T/T_{F,i}$ is obtained by fitting time-of-flight images to finite-temperature Fermi-Dirac distributions. 
Note that at the end of the evaporation the populations of the additional chromium minority components Cr$|i\!>1\rangle$, initially loaded within the BODT, are negligible, owing to the combined effect of inelastic two-body losses occurring throughout the evaporation ramp, and to the lack of thermalization with lithium, given the spin-selective character of the ``Feshbach cooling'' mechanism.   

The evaporation trajectories summarized in Fig.~\ref{Fig3}(a), which we experimentally found to be optimal to realize large and degenerate Li-Cr Fermi mixtures, can be also adapted to produce single-species samples of either species. For lithium, this is straightforward: Without loading the chromium component, the same BODT power ramps above discussed yield crossover superfluids of more than 4.5$\times$10$^5$ pairs, when the bias field is tuned towards the pole of the broad intra-species Feshbach resonance at 832 G. Such a number can be further increased up to about 6$\times$10$^5$ when the same time evolution of the trap depth is realized by means of the sole IR light of the BODT, our setup featuring performances similar to, and even slightly better than, those reported in Ref. \cite{Burchianti2014}.

To realize polarized Fermi gases of $^{53}$Cr, the protocols above discussed can be modified only partially, owing to the fact that quantum degeneracy of this species can so far be obtained only via sympathetic cooling with lithium.
For this reason, reducing the  Li number initially loaded into the BODT is not beneficial. Yet, a slight increase of the green-to-IR power ratio during the evaporation allows us to obtain about 70$\%$ larger Cr samples at 220(20) nK, at the expense of a significant reduction of both Li components, which can be eventually completely removed at the end of the evaporation stage by further increasing the power of the green BODT arm.
Sympathetic cooling of $^{53}$Cr with $^6$Li thus appears as a promising route to realize large Fermi gases of this yet poorly explored atomic species, so far produced only in combination with its most abundant bosonic isotope $^{52}$Cr \cite{Naylor2015}. In fact, the possibility to exploit the different Li and Cr polarizabilities to the IR and green lights of our BODT, absent when isotopic Cr mixtures are considered, together with our substantially larger Cr MOT, allows for an almost 200 fold increase in the $^{53}$Cr atom number which can be brought to $T/T_{F, Cr}\!\lesssim$1, relative to previous studies \cite{Naylor2015}.  

\section{Increasing quantum degeneracy in a crossed BODT}
\label{crossed}
\begin{figure*}[t!]
\begin{center}
\includegraphics[width=178mm]{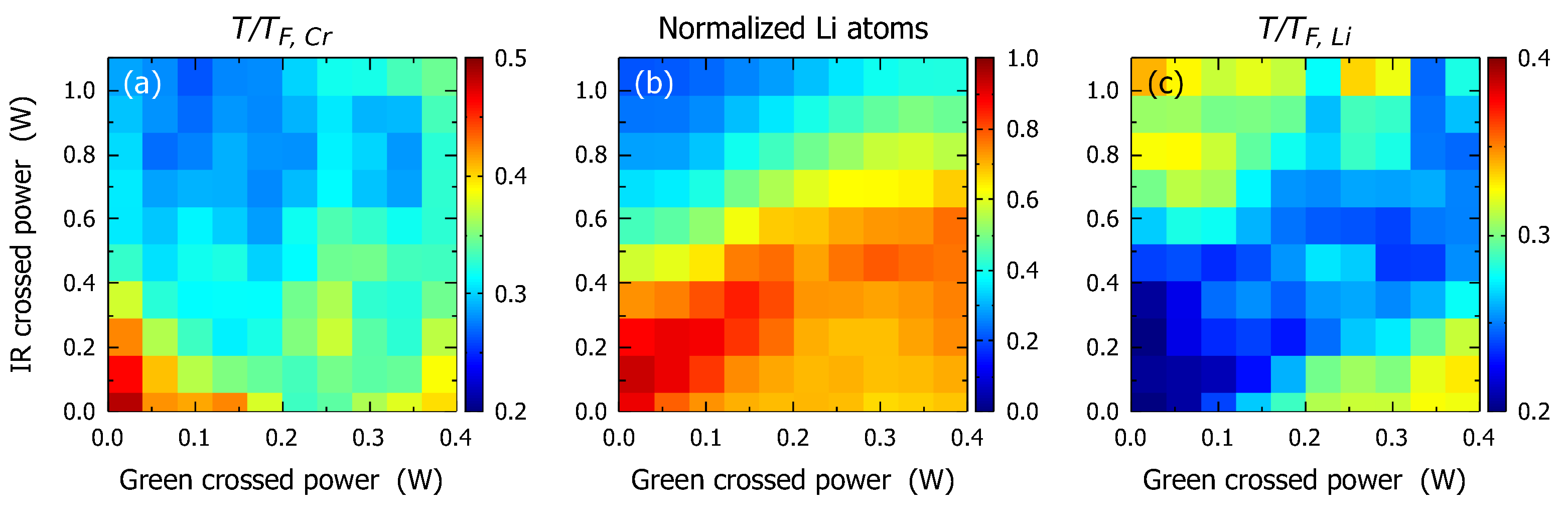}
\caption{
\textbf{(a)} Normalized chromium temperature, $T/T_{F, Cr}$, monitored as a function of green and IR powers of the crossed BODT. The degree of degeneracy is obtained as the average value extracted from fitting at least four independent Cr images, acquired at 4.6 ms time-of-flight. 
\textbf{(b)} Lithium atom number, normalized to its value measured in the sole main BODT at the end of the evaporation ramp, as a function of green and IR powers of the crossed BODT. 
\textbf{(c)} Same as in panel (a) for the reduced lithium temperature $T/T_{F, Li}$. For lithium, a 3 ms time-of-flight expansion was employed.}  
\label{Fig4}
\end{center}
\vspace*{-5pt}
\end{figure*}
As discussed in the previous section, the degree of degeneracy, obtained at the end of the evaporation stage discussed therein, is quite different for the two mixture components, with lithium being highly degenerate while chromium featuring $T/T_{F, Cr}\!\sim$0.5.
Further decreasing the BODT trap depth does not lead to any substantial gain in phase-space density, a reduced temperature being counterbalanced by a drop in the atomic densities for both species.

To overcome this problem, once the evaporation stage is ended, we raise a second bichromatic trapping beam, which crosses the main BODT at an angle of about 15 degrees from the vertical direction. Such a secondary beam is obtained by exploiting the same laser sources, recycling part of the IR and green powers of the main BODT, damped at the end of the evaporation procedure. 
Both IR and Green crossed beams are almost circular, and at the atom position they feature waists of about 60 $\text{µm}$ and 70 $\text{µm}$, respectively.
Since such a secondary trap is oriented almost vertically, it does not substantially modify the trap depth experienced by the two atomic components, that is  set only by the  main BODT beam. On the other hand, the crossed beam allows us to controllably tune the confinement of the two atomic clouds in the horizontal plane, and primarily along the weakest axial direction of the main BODT, where the Li and Cr trap frequencies, at the end of the evaporation, are otherwise solely set by our moderate magnetic-field curvature.
Therefore, adiabatically loading the ultracold mixture into the crossed BODT may lead to a substantial enhancement of the atomic density, hence of the Fermi energy, while not affecting the temperature of the two components.

Note that, also in this case, lithium and chromium are affected differently by the IR and green components of the crossed beam: While chromium is confined by both lights, the lithium sample is tightly confined by the IR, and it is almost symmetrically anti-confined by the green light.
As a result, the crossed BODT represents an appealing tool that allows us to modify the Li and Cr density distributions almost independently, and to controllably tune the lithium-to-chromium density ratio.

In order to test this possibility, once the evaporation ramp is performed, we raise up the crossed BODT at various IR and green power levels through a 50 ms linear ramp. After about 50 ms, we then record time-of-flight images of both Li$|1\rangle$ and Cr$|1\rangle$, and obtain the corresponding atom number and degree of degeneracy by fitting the atomic clouds to a finite-temperature Fermi-Dirac distribution.
The results of this characterization are summarized in the contour plots in Figure~\ref{Fig4}: Panels (a) and (c) show the chromium and lithium normalized  temperatures $T/T_{F, Cr}$ and $T/T_{F, Li}$, respectively, as a function of green and IR powers of the crossed beam. 
Fig.~\ref{Fig4}(b) presents the corresponding trend for the Li$|1\rangle$ atom number, normalized to 3.5$\times$10$^5$, which is the value obtained without application of the crossed beam.
The chromium component, characterized by an axial size in the main BODT roughly two times smaller than the lithium one, is efficiently transferred into the crossed trap at all green and IR powers herein explored, resulting in a Cr number, not shown in Fig.~\ref{Fig4}, that varies less than 15$\%$ throughout the investigated parameter space.   

One can notice the qualitatively different response of the two mixture components to the crossed BODT.
For chromium, application of either crossed beam leads to a substantial increase in the degree of degeneracy: As shown in Fig.~\ref{Fig4}(a), several IR and green combinations yield a two-fold decrease of $T/T_{F, Cr}$, passing from about 0.5 down to 0.25, solely caused by the large Cr density increase within the crossed trap. 

Lithium, experiencing opposite IR and green potentials, exhibits quite different trends, with respect to the chromium component.
As shown in Fig.~\ref{Fig4}(b), the atomic fraction which can be collected in the crossed BODT exhibits a sizable drop, both with increasing IR and green power.
Owing to the 3-fold higher polarizability at IR light of Li relative to Cr, and given an axial size of the lithium Fermi gas in the single-beam BODT about two times larger than the one of chromium, a relatively low IR power level already suffices to produce a tightly-confining potential, within which only a comparably small fraction of the initial lithium atoms can be collected and stored. As a result, the corresponding $T/T_{F, Li}$ is increased, see Fig.~\ref{Fig4}(c).
Application of the sole green crossed beam has a qualitatively similar effect, arising however from additional anti-confinement, rather than confinement as for the case of the IR beam. 
For low powers, although the Li number in the crossing region is only moderately reduced, see Fig.~\ref{Fig4}(b), the central density is sizably depleted, resulting in an increased $T/T_{F, Li}$ in Fig.~\ref{Fig4}(c). 
An even higher green power, not shown in the plots, causes a strong perturbation of the main BODT potential, and a substantial drop of the Li atomic density in the crossing region. 
On the other hand, when both lights are applied, over a quite wide range of parameters we observe an efficient storage of lithium atoms in the crossed BODT, not  significantly increasing $T/T_{F, Li}$, see Fig.~\ref{Fig4}(c). As expected, this occurs roughly around the diagonal of Figs.~\ref{Fig4}(b) and (c), where the anti-confinement of the green beam is (more than) counterbalanced by the IR light.

\begin{figure}[t!]
\begin{center}
\includegraphics[width=80mm]{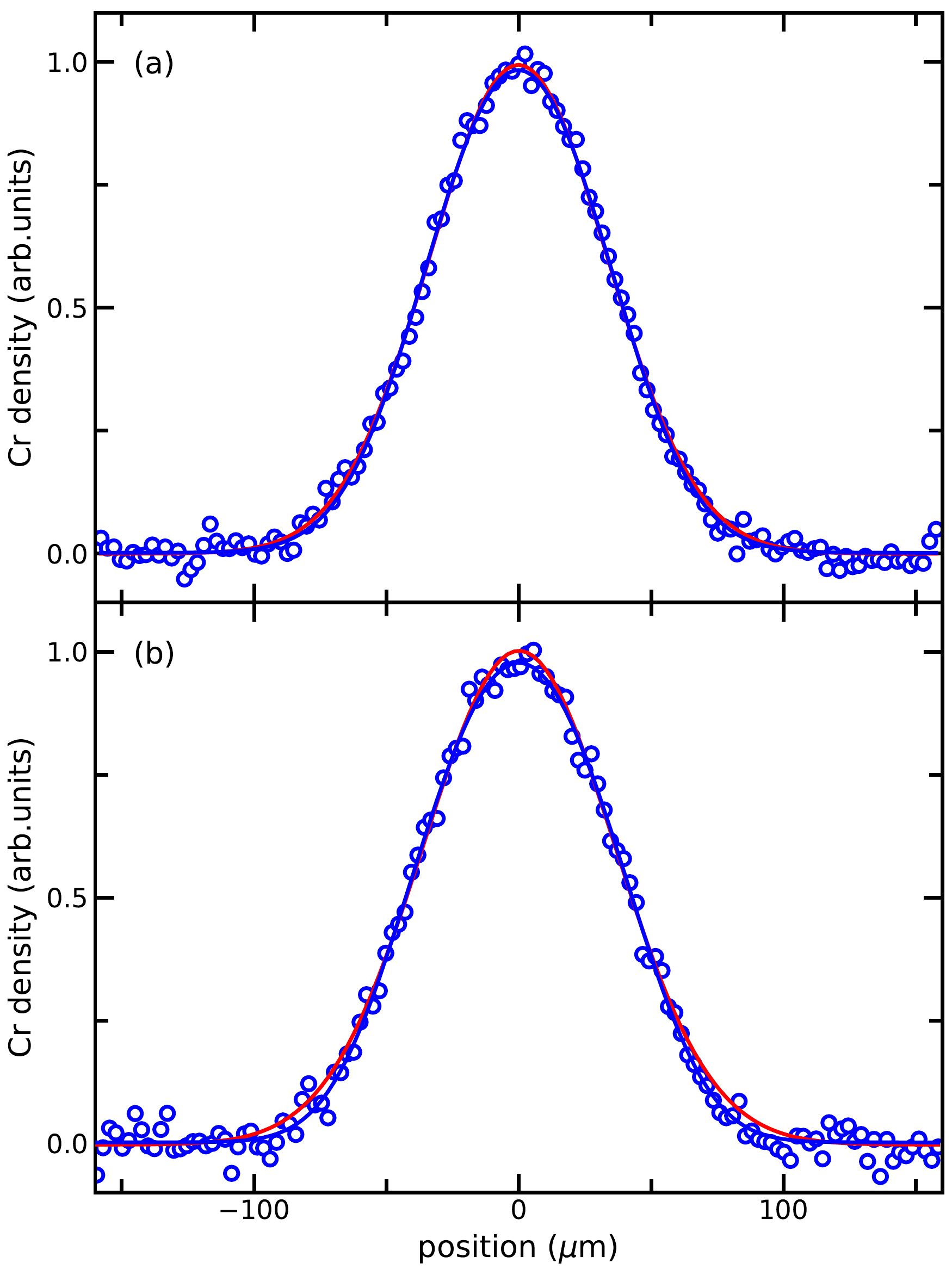}
\caption{
\textbf{(a)} Axially-integrated density profile of a chromium gas (blue circles), confined into the main BODT trap only, obtained from the average of 20 independent images recorded after 5.0 ms of time of flight. 
Experimental data are compared with best fits to a Fermi-Dirac (blue line) and a Gaussian (red line) distribution. From the former fit we obtain $T/T_{F, Cr}$=0.45(7).
\textbf{(b)} Same as panel (a) when the chromium component is loaded in the crossed BODT, with IR and green powers being set to 0.77 W and 0.08 W, respectively. The fit to a Fermi-Dirac distribution yields in this case $T/T_{F, Cr}$=0.25(2).}  
\label{Fig5}
\end{center}
\vspace*{-10pt}
\end{figure}

These observations highlight how the crossed BODT significantly enhances the parameter space which can be explored with the Li-Cr mixture in our setup: By simply tuning the (absolute and relative) powers of the two crossed lights, one can pass from the regime where lithium is highly degenerate and chromium is an almost thermal gas, to the opposite one. 
Most importantly, over a sizable range of parameters we can simultaneously achieve a high degree of degeneracy for both $^6$Li and $^{53}$Cr components.
As an example, in Figure~\ref{Fig5} we show radially-integrated density profiles of a chromium Fermi gas, imaged after 5.0 ms of time-of-flight expansion from the sole main BODT trap (panel (a)) and from a crossed BODT combining 0.77 W IR and 0.08 W green light (panel (b)). Data are compared with best fits to a Gaussian and to a Fermi-Dirac distribution function, shown in red and blue, respectively. 
In both trap configurations, the lithium sample features a roughly constant and low $T/T_{F, Li}$ value, see Fig.~\ref{Fig4}(c).
For the chromium component, application of the crossed beam negligibly affects the atom number, constantly about 1.0$\times$10$^5$, while it substantially increases the sample degree of degeneracy, passing from $T/T_{F, Cr}$=0.45(7) in the sole main BODT beam, Fig.~\ref{Fig5}(a), to $T/T_{F, Cr}$=0.25(3) when Cr is loaded into the crossed bichromatic trap.

\section{Conclusions}
\label{Conclusions}  
In conclusion, we have reported on the experimental procedures that we devised to realize large and degenerate Fermi mixtures of lithium and chromium atoms.
The present work, combined with the recent discovery and characterization of interspecies Li-Cr Feshbach resonances \cite{Ciamei2022B}, makes $^6$Li-$^{53}$Cr mixtures an appealing playground with which to investigate a variety of yet unexplored phenomena, spanning from the few- and many-body physics of strongly-interacting fermionic matter in presence of a large mass asymmetry, to the formation of a new kind of paramagnetic polar molecules in the ultracold regime \cite{Zaremba2022}.

Owing to the peculiar chromium-lithium mass ratio, extremely close to the critical values above which non-Efimovian three- and four-body cluster states are predicted to emerge \cite{Kartavtsev2007,Endo2011,Endo2012,Blume2012A,Bazak2017}, our mixture may provide an exemplary benchmark for a wealth of theoretical predictions, lacking experimental observation in any physical system, so far. Moreover, the collisional stability predicted for such exotic few-body states \cite{Levinsen2011, Endo2016, Bazak2017} makes them appealing also from a many-body perspective.
For instance, novel types of quasi-particles could emerge in the \textit{light} impurity problem: Besides Fermi polarons and dressed dimers \cite{Massignan2014}, lithium impurities embedded in a degenerate Fermi gas of chromium may indeed exhibit more complex quasi-particles \cite{Mathy2011,Endo2016, Liu2022B}, connected with the existence of higher-order few-body cluster states in the vacuum. 
In this respect, the ability to widely tune both the degree of degeneracy and the relative densities of the Li and Cr components in our crossed BODT offers a compelling opportunity to investigate, within the same physical setup, both heavy and light impurity problems within fermionic media. 

Furthermore, our doubly-degenerate mixture also appears optimally-suited to investigate both $s$- and $p$-wave pairing, and the possible emergence of superfluid states, close to Li-Cr Feshbach resonances already identified \cite{Ciamei2022B}. 
The availability of large Li and Cr ultracold atomic samples, and the tunability of their densities enabled by our crossed bichromatic potential, provides a promising system to realize high phase-space density samples of bosonic Feshbach dimers  \cite{Duda2021, Lam2022}, also in light of their increased collisional stability under strongly interacting conditions \cite{Petrov2004,Jag2016}, when compared to those realized from Fermi-Bose or Bose-Bose mixtures. 
The realization of a Bose-Einstein condensate of LiCr Feshbach dimers would not only pave the way to the exploration of resonant superfluidity within a mass-imbalanced Fermi mixture, but it would also represent an optimal starting point to realize,  through optical transfer schemes to deeply-bound levels, degenerate Bose gases of paramagnetic polar molecules \cite{Zaremba2022}.

Finally, the experimental protocols described in this work are appealing also to realize degenerate Fermi gases, or spin-mixtures, of $^{53}$Cr atoms: Either via sympathetic cooling with lithium herein discussed or, eventually, by adapting our strategies to single-species setups.
This fermionic species, yet almost unexplored, represents an interesting system on its own, as it combines a sizable magnetic dipole moment, comparable with the one of erbium, with a wide tunability of intra-species interaction, enabled by wide and isolated Feshbach resonances, predicted for its two lowest Zeeman states \cite{Pavlovic2011,Simoni2022}. As such, $^{53}$Cr binary spin-mixtures may offer the compelling opportunity of investigating the physics of the BCS-BEC crossover in the presence of weak dipolar interactions. 

\begin{acknowledgments}
We thank  D. Petrov, B. Laburthe-Tolra, A. Simoni and the LENS Quantum Gases group for useful discussions. This work was supported by the ERC through grant no.\:637738 PoLiChroM, by the Italian MIUR through the FARE grant no.~R168HMHFYM P-HeLiCS, and through the EU H2020 Marie Sk\l{}odowska-Curie program (fellowship to A.C.).
\end{acknowledgments}

\vspace*{0pt}

%

\end{document}